\documentstyle[12pt]{article}
\input epsf
\begin{document}

\newcommand{\pho}{\tilde{\gamma}}
\newcommand{\gl}{\tilde{g}}
\newcommand{\sneu}{\tilde{\nu}}
\newcommand{\st}{\tilde{t}}
\newcommand{\sq}{\tilde{q}}
\newcommand{\se}{\tilde{e}}
\newcommand{\ch}{\chi^{\pm}}
\newcommand{\neut}{\chi^{0}}
\newcommand{\gsi}{\,\raisebox{-0.13cm}{$\stackrel{\textstyle>}
{\textstyle\sim}$}\,}
\newcommand{\lsi}{\,\raisebox{-0.13cm}{$\stackrel{\textstyle<}
{\textstyle\sim}$}\,}
\newcommand{\be}{\begin{equation}} \newcommand{\ee}{\end {equation}}

\rightline{RU-96-71} 
\rightline{hep-ph/yymmddd}
\rightline{August 2, 1996} 
\baselineskip=18pt \vskip 0.7in 
\begin{center} {\bf \LARGE Phenomenology of Charginos and Neutralinos in the
Light Gaugino Scenario}\\ 
\vspace*{0.9in} 
{\large Glennys R. Farrar}\footnote{Research supported in part by
NSF-PHY-94-23002} \\ 
\vspace{.1in} 
{\it Department of Physics and Astronomy \\ Rutgers
University, Piscataway, NJ 08855, USA}\\ 
\end{center} 
\vspace*{0.2in}
\vskip 0.3in 
{\bf Abstract:} The light gaugino scenario predicts that the lighter
chargino mass is less than $m_W$, gluino and lightest neutralino masses
are $ \lsi 1$ GeV, and the dominant decay mode of charginos and non-LSP 
neutralinos is generically to three jets.  The excess "$4j$" events 
observed by ALEPH in $e^+ e^-$ annihilation at 133 GeV may be evidence 
that $m(\ch_1) = 53$ GeV.  If so, $m(\ch_2) = 110-121$ GeV,
$m(\neut_2) = 38-63$ GeV, $m(\neut_3) = 75-68$ GeV; $m(\sneu_e)$ is 
probably $\sim m(\ch_1)$.  A detailed
analysis of the multi-jet events is needed to exclude this possibility. 
Consequences for FNAL and higher energy LEP running are given.   

\thispagestyle{empty} 
\newpage 
\addtocounter{page}{-1}
\newpage

In the light gaugino scenario the lighter chargino mass is $\le m_W$.
As we shall see, charginos and neutralinos generally decay to 3 jets,
and squarks to two jets.  The missing energy is too small in both
cases to be a useful signature.  The purpose of this paper is twofold:
(i) to define the phenomenology of charginos and neutralinos in this
scenario, and (ii) to investigate the possibility that the ALEPH
excess ``4 jet'' events in 133 GeV $e^+ e^-$ annihilation come from decay
of the lighter chargino.  

Some supersymmetry (SUSY) breaking scenarios produce negligible
tree-level gaugino masses and scalar trilinear couplings\cite{bkn}
($M_1=M_2=M_3=A=0$).  As pointed out in refs. \cite{f:99,f:101}, this has
several attractive theoretical consequences such as the absence of the ``SUSY
CP problem" and avoidance of certain cosmological problems.  Although massless
at tree level, gauginos get calculable masses through radiative 
corrections from electroweak (gaugino/higgsino-Higgs/gauge boson) and
top-stop loops. Evaluating these within the constrained parameter
space leads to a gluino mass range $m_{\tilde{g}}\sim \frac{1}{10} -
\frac{1}{2}$ GeV and photino mass range $m_{\tilde{\gamma}} \sim
\frac{1}{10} - 1 \frac{1}{2}$ GeV\cite{f:99,f:101}.  The chargino and other
neutralino masses are functions only of $\mu$ and $tan \beta$.  In
particular,  
\begin{equation}
2 M_{\ch}^2 = \mu^2 + 2 m_W^2 \pm \sqrt{\mu^4 + 4 m_W^4 cos^2
2 \beta + 4 m_W^2 \mu^2}, 
\label{mch}
\end{equation}
so one chargino is lighter, and the other heavier, than $m_W$.  The
photino is an attractive dark matter candidate, with a correct abundance
for parameters in the predicted ranges\cite{f:100}. 

Due to the non-negligible mass of the photino compared to the
lightest gluino-containing hadron, prompt photinos\cite{f:24} are not
a useful signature for the light gluinos and the energy they
carry\cite{f:51}. Gluino masses less than about $ \frac{1}{2}$ GeV are
largely unconstrained\cite{f:95}\footnote{The recent
claim\cite{murayama:4j} that LEP $Z^0 \rightarrow 4j$ data can be 
used to exclude light gluinos is premature.  Although the statistical
power of the data is sufficient, the relevant angular distributions
are particularly sensitive to higher order effects such as 5-jet
production.  This is evidenced by the spread of predictions in the absence
of light gluinos shown in Fig. 5 of ref. \cite{opal4j1}, which is an
order of magnitude larger than the effect due to light gluinos (see
Fig. 2 of ref. \cite{f:82}).  Therefore as stressed in \cite{f:82}, no
conclusion can be drawn until the NLO calculation of the $Z^0 \rightarrow
4j$ matrix element is available.}, although such light gluinos would cause
modifications in jet distributions\cite{f:82,clav:jets,bern:gljets} and have
other indirect effects, e.g., on the running of $\alpha_s$\cite{f:32,jk}.
The lifetime of the gluon-gluino bound state ($R^0$) is predicted to be
$10^{-5}-10^{-10}$ sec\cite{f:101}.  Proposals for direct searches for
hadrons containing gluinos, via their decays in $K^0$ beams and
otherwise, are given in Ref. \cite{f:104}. 

For the purposes of detecting squarks and charginos, the crucial
phenomenological difference arising when the gluino is light rather
than heavy as is usually assumed, is that the gluino is long-lived.
Therefore it makes a jet rather than missing energy\cite{f:51}.
Squarks decay to gluino and quark, thus generating two jets with
negligible missing energy\footnote{The missing energy associated with
the ultimate decay of the lightest $R$-hadron to a photino is far
below $E_{miss}$ cuts in squark searches\cite{f:105}.}.  QCD
background makes it impossible, with present jet resolution, to search
at hadron colliders in the dijet channel for masses lower than about 200
GeV; a search for equal-mass dijet pairs as suggested in \cite{f:105}
has not yet been completed.  At present, the best squark mass limits
come from the hadronic width of the $Z^0$ and are only $\sim 50-60$
GeV\cite{clavsqlim,bhatsqlim}.
 
The rate for a two-body decay of a chargino to an SU(2) doublet can
be written  
\be 
\Gamma(\chi^+ \rightarrow \tilde{\bar{D}} + U) = N_c cos^2\phi \frac{\alpha_2
|k|}{8 M_{\ch}^2} (M_{\ch}^2 + m_q^2 - M_{\tilde{q}}^2),
\label{Gch2} 
\ee 
where $N_c$ is the QCD multiplicity of the fermion (3 for quarks, 1
for leptons), $|k|$ is the CM momentum of the final state particles,  
and $cos \phi$ is the effective amplitude that the chargino is a
$\tilde{w}^{\pm}$.\footnote{Here, $cos \phi = U_{i1}$ in the notation
of \cite{gh};  for $\Gamma(\chi_i^+ \rightarrow \tilde{U} + \bar{D})$, $cos
\phi =  V_{i1}$.  In eq. (\ref{Gch3}), $cos \bar{\phi} \equiv (U_{i1} +
V_{i1})/2$.}  For $\mu,~tan \beta$ values in the relevant range (see below),
charginos are an approximately equal admixture of wino and higgsino.  Thus
$\chi_i^+ \rightarrow \tilde{t_R} + \bar{b_L}$ is an important decay mode,
if kinematically allowed.

The total rate for hadronic three-body decays is proportional to the decay rate
calculated long ago for the gluino\cite{hk:taugluino}, so modifying
the overall factor appropriately: 
\be
\Gamma(\chi^+ \rightarrow \gl + \bar{D} + U) =  \frac{\alpha_s \alpha_2 cos^2
\bar{\phi}}{16 \pi}  \frac{M_{\ch}^5}{M_{\sq}^4},
\label{Gch3} 
\ee 
where for brevity squark masses have been taken equal and quark and gluino
masses neglected.

The most important feature to note is that if the gluino is light and
$M_{\sq}$ is not $ >> m_W$, the overwhelmingly dominant three-body
final state is $q \bar{q'} \gl$, produced by a virtual
squark\footnote{For $M_{\sq} \sim m_W$, it is a factor $\frac{2 \cdot
8 \alpha_3}{(2 \cdot 3 + 3) \alpha_2} \sim 5$ larger than virtual $W$
decay, summing over the two generations of light quarks and three 
generations of leptons, and a factor $8\frac{\alpha_3}{\alpha_2} \sim
25$ larger than virtual slepton or sneutrino decay.  The former gives 
$\neut q \bar{q'}$ two-thirds of the time and otherwise $\neut l \nu$,
while the latter gives $l \nu \neut$.}.  Thus unless two-body decays
are possible, or sneutrinos or sleptons are much lighter than squarks,
or sfermions are much heavier than $W$'s, the lighter chargino and the
three heavier neutralinos will almost always decay to three jets with no
missing energy.  Since $M(\ch_2) \ge m_W$, the two-body decay $\ch_2
\rightarrow W^{\pm} + \pho$ is allowed and will be an important decay
mode, except very near $M(\ch_2) = m_W$.  Another possible exception to the
dominance of $q \bar{q} \gl$ final states occurs when $tan \beta$ is close
to one: in this case,  $\neut_3$ has very little $\tilde{z}$ component and 
its dominant decay mode may be through a stop-top loop to $\gamma \pho$. 

If $M_{\sq} \sim M_{\ch}$, the matrix element favors one jet being
much softer than the other two.  This is easy to see for the two-body
decay $\ch \rightarrow \sq + q'$ with $M_{\ch} >> M_{\ch} - M_{\sq} >0$, for
which the primary jet ($q'$) has energy $\approx M_{\ch}-M_{\sq}$ in
the $\ch$ rest frame, while the quark and gluino from the squark decay
each have $E \approx M_{\sq}/2$.  If one jet is much softer than the
other two, particles from that jet will often be merged into the hard
jets by the jet finding algorithm and a chargino will give rise to
``$2j$'', i.e., a multijet system designated as two jets.  With a
sample of candidate chargino events, analysis of the relative 
number of events with resolved jets can give limits on the
intermediate squark mass: $M_{\sq} >> M_{\ch}$ leads to final states
with jets of comparable energy, while $M_{\sq} \sim M_{\ch}$ leads to
states in which one jet has much lower energy than the other two and
thus is less easily resolved. 

In the LEP 133 GeV run, ALEPH observed 16 "$4j$" events when 8.6 were
expected \cite{aleph:4j}.  The excess was localized in the
total-dijet-mass range 102-108 GeV, where 9 events were observed when
0.8 was expected.  Moreover most of the events in this peak region are
not characteristic of the SM expectation with regard to their angular
distribution and dijet charge-difference.  This reduces the likelihood that
the events were simply a statistical fluctuation of a standard model process.

Assuming this excess is due to pair production of equal mass particles which
decay to two jets\footnote{E.g., $h~A$, although that is unlikely to be the 
origin of the excess events, since its cross section is 0.49 pb for 
$M(h) = M(A) = 53$ GeV and there is no observed excess of $b \bar{b}$'s in 
the events.} implies a cross section of $3.1~\pm~1.7$ pb.  Taking the 4 LEP 
experiments together gives a cross section of $1.2 ~\pm ~0.4$
pb\footnote{Results from the ensemble of LEP experiments are taken from P. 
Mattig, XXVIII Intl. Conf. for High Energy Physics, Warsaw, July 24-31, 
1996.}. With this cross section it is not particularly improbable that one of
the 4 experiments should get 16 events (SM model expectation 8.6) out of
the total of 49 events observed, and the others together have 33 (with 26.4
expected in the SM).  Furthermore in the 102-108 GeV total-dijet-mass region
the other experiments also observed an excess: 6 events when 2.6 were expected.

Pair production of $\sim 53$ GeV charginos could give rise to ``$4j$''
events with total dijet mass of 105 GeV at the observed rate.  The ALEPH
jet-reconstruction algorithm explicitly merged $5j$ to $4j$, and ignored 
the small number of clear $6j$ events.  An important point is that due to the 
experimental imprecision in energy measurements, LEP experiments rescale the 
momenta and energies of jets to enforce overall energy momentum conservation.  
The directions of the jets are assumed to be accurately measured, 
and the jets taken to be massless.  Then, since energy-momentum conservation
provides 4 equations, up to 4 jets can  be independently rescaled.  However
if the event actually contains 6 jets and the invariant mass of merged pairs
of jets is non-negligible, the procedure of rescaling the magnitudes of the
momenta of the 4 jets distorts the invariant masses of the "dijets".  Detailed
Monte Carlo study is needed to assess the extent to which this causes the
chargino or neutralino invariant mass peaks to be lost, and the beam
energy dependence of this effect.

The chargino production cross section depends on $\mu,~tan \beta$ and
$M_{\sneu_e}$.  $M_{\ch_1}= 53$ GeV requires a relation between
$\mu$ and $tan \beta$, eq. (\ref{mch}), which also ensures that the chargino
is an approximately equal mixture of higgsino and wino.  Therefore the main
uncertainty in cross section is due to its sensitivity to the electron
sneutrino's mass, $M_{\sneu_e}$.  Imposing $M_{\neut_2} \ge 38$ GeV (see 
below), causes $[\mu,tan\beta]$ to range between $[45,1.6]$ and $[70,1]
$\footnote{We
take $tan \beta \ge 1$ without loss of generality because in the absence
of tree-level gaugino masses and scalar trilinear couplings, the chargino
and neutralino spectrum is unchanged by $tan \beta \rightarrow 1/(tan \beta)$;
only the roles of the higgsinos, $h_U$ and $h_D$, are interchanged in the
eigenstates.}.  For instance for $M_{\sneu_e} = 60$ GeV, $tan \beta = 1.4$
and $\mu = 56$ GeV, one finds $\sigma(\chi^+ \chi^-) = 3.6$ pb at $E = 133$
GeV, and 2.1 (1.9) pb at $E=161~(190)$ GeV.  The smallest possible $\chi^+_1
\chi^-_1$ cross section at 133 GeV is 2.4 pb, for $M_{\sneu_e} = 50$ 
GeV, while the largest possible cross section is 14 pb (for $M_{\sneu_e}>>
M_{\ch_1}$, $tan \beta = 1$ and $\mu = 68.4$ GeV).  

If at 133 GeV most $\ch_1 \rightarrow q \bar{q} \gl$ decays are captured
by the ALEPH analysis procedure, a cross section to "$4j$" as low as 1.2
pb suggests a competing decay mode.  This could be $\ch_1 \rightarrow l \sneu$,
with the $\sneu$ decaying to lsp and neutrino and thus going undetected.
The branching fraction for $\ch_1 \rightarrow l \sneu$ is a very sensitive
function of the sneutrino mass and also depends on the squark mass and the
number of light sneutrinos.  The mass splitting must be less than about $0.5$
GeV in order not to excessively reduce the branching fraction to the hadronic
channel. With such a small splitting, the lepton energy is too low for events
with two leptonic decays to be accepted.  Since the detection efficiency for
leptons drops very rapidly with momentum, events with one chargino
decaying hadronically to $q \bar{q} \gl$ and the other to lepton and
missing energy may be difficult to detect.  The LEP experiments should
explore what limits can be placed on this possibility.

We can hope that higher integrated luminosity will allow the 53 GeV
chargino hypothesis to be promptly confirmed or excluded.  With greater CM
energy the jet systems from each chargino decay will be better
collimated and more readily separated from one another, so the angular
distribution of the jet systems can be more cleanly determined than at
lower energy.  It should $\sim 1 + cos^2 \theta$ because charginos are
spin-1/2.  On the other hand, Monte Carlo simulations\footnote{M.
Schmitt, private communication.} show that the resolution in dijet mass
difference does not improve significantly with energy even for genuine 
two-body decays of pair-produced particles, due to reduced effectiveness
of the energy-momentum constraint at higher energy.  This will probably be
an even more severe problem for the $6j$ case at hand.  It is ironic that  
$W^+ W^-$ production will be a non-trivial background at 161 GeV and above, 
since they decay with rather high probability to $> 2j$ (40 \% of
$Z^0$ decays contain $\ge 3j$, with $y_{cut} = 0.01$) and their cross
section is about a factor of 5 larger than that of charginos, whose
cross section at 161 GeV, extrapolating from 1.2 pb at 133 GeV,
is 0.8 pb.  As of the Warsaw conference, with each experiment having about 
3.2 pb$^{-1}$ integrated luminosity at 161 GeV, no excess of events was 
observed.  This gives a 95\% cl upper limit of 0.85 pb on the production of 
4j events.  However the effficiency of seeing 6j events with the 4j
analysis is presently unknown, so this limit cannot be directly applied
to $\chi^+ \chi^-$ production.

What else can be said about the SUSY spectrum under the 53 GeV
chargino hypothesis?  First of all, if a sneutrino (selectron) were
light enough that the two-body decay $\ch \rightarrow \sneu + l$ or 
$\ch \rightarrow \tilde{l} + \nu$ (branching fraction $b_l$, taken together) 
could compete with the hadronic decays (branching fraction $b_h$), one should 
see other signatures.  Events with two leptons and missing
energy should occur at a rate $(b_l/b_h)^2$ compared to the ``$4j$''
events, and there should be mixed final states with ``$2j$'', a single
lepton, and missing energy at a relative rate $2 b_l/b_h$.  The
charged lepton will be soft or hard, depending on which decay ($\ch
\rightarrow \sneu + l$ or $\ch \rightarrow \tilde{l} + \nu$) dominates.  
These final states would have shown up in LEP SUSY searches unless the charged
lepton is extremely soft, so we can conclude that $~m(\tilde{l}) > 53$ GeV and
$m(\sneu) \gsi 50$ GeV\footnote{The 3 GeV mass difference required to see the 
soft leptons is evident in Fig. 7 of \cite{aleph:neut96}.}.

We can also deduce that the right-stop is probably heavier than the
chargino.  A stop lighter than the top would decay through
FCNC mixing, to a gluino and presumably charm quark\footnote{Information on
FCNC mixing involving the third generation is poor, so mixing between
third and first generations may also occur.}. If the decay
$\ch_1 \rightarrow \st_1 + b$ were allowed, each event would contain soft $b
\bar{b}$ and hard $ c \bar{c}$ jets in addition to the gluino jets.
However even if $\ch_1 \rightarrow \st_1 + b$ is not allowed, when chargino
decay is mediated by a real or virtual stop we would expect two hard
charm-quark jets in each chargino event. ALEPH has searched for $b$, $
\bar{b}$, $c$, and $ \bar{c}$ jets in their $4j$ sample.  They found
only a single event consistent with a displaced vertex or excess
lepton activity from the $e$ or $\mu$ produced in 20\% of the $c$ and 
$b$ decays, and no evidence for an excess in comparison to QCD
expectations\cite{aleph:4j}.  So we conclude that $M_{\st_1}$ is probably
$ \gsi 53$
GeV\footnote{Chargino decay through virtual stops may be highly
suppressed by the FCNC factor, which is irrelevant for on-shell stops,
so for $M_{\st_1} \gsi 53$ GeV we cannot use such reasoning to decide
how $M_{\st}$ compares to the other squark masses.}. 

Now let us turn to the neutralinos and heavier chargino.  In the
tree-level massless gaugino scenario, the lightest neutralino gets its
mass through radiative corrections\cite{f:96,pierce_papa} and has mass
$\lsi 1 \frac{1}{2}$ GeV\cite{f:99,f:101}.  It is practically pure
photino.  The masses of the others are determined by $\mu$ and $tan
\beta$, since tree-level gaugino masses vanish.  Since $\neut_1
\neut_1$ and $\neut_1 \neut_2$ production is severely suppressed by
the absence of a higgsino component in $\neut_1$, the only constraints
of importance come from $\neut_2 \neut_2$ production.  Fixing the
relation between $\mu$ and $tan \beta$ so $M(\ch_1) = 53$ GeV, implies
$M(\neut_2) \ge 45$ GeV if $\mu \ge 50$ GeV.  In this case $\neut_2$
could not be pair-produced in $Z^0$ decays.  A somewhat lower mass
$\neut_2$ is actually not excluded, since $\neut_2$ decays practically
exclusively to $3j$ final states and therefore would not have been
detected in existing neutralino searches.  Allowing $M(\neut_2)$ = 38
GeV (i.e., $\mu = 45$ GeV, $tan \beta = 1.6$) would result in an
increment of about 10 MeV in the hadronic width of the $Z^0$, whose
PDG value is $\Gamma_{had}(Z^0) = 1741 \pm 6$ MeV.  For reference, the
SM prediction for $\Gamma_{had}(Z^0)$ would drop by about this amount
if $\alpha_s(M_Z)$ were in fact $0.115$ as suggested by low energy
data\footnote{In the light gluino scenario, however, evolution from
$\alpha_s=0.11 $ at 10 GeV leads to $\alpha_s(M_{Z^0}) = 0.122$, in
agreement with the quoted LEP average of 0.125.}.  For $M_{\neut_2} =
38$ GeV, $\neut_2$ has a large $\tilde{h}_D$ component ($\ch_2 = -0.86
\tilde{h}_D -0.39 \tilde{h}_U + 0.31 \tilde{z}^0$).  This leads to an
excess in the $b \bar{b} \gl$ final state in $\neut_2$ decay, but not
enough to affect $R_b$ significantly. 

\begin{figure}
\epsfxsize=\hsize
\epsffile{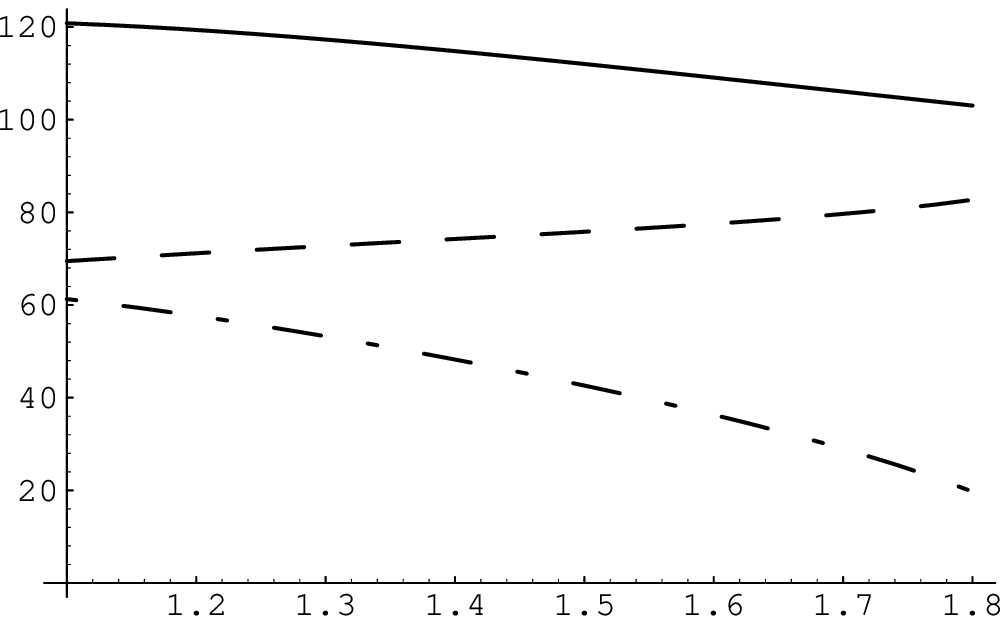}
\caption{Masses in GeV of the heavier chargino and second and third
neutralinos as a function of $tan \beta$, fixing the lighter chargino
mass to 53 GeV (solid, dash-dot, and dash curves, respectively).}  
\label{masses}
\end{figure}

Having restricted the range of $[\mu, tan \beta]$ to $[45,1.6] -
[70,1]$, we find the heavier {\it ino} mass ranges: $M(\neut_3) =
76-68$ GeV, $M(\neut_4) = 118-132$ GeV and $M(\ch_2) = 110 - 120$ GeV.
See Fig. \ref{masses}.

At 133 GeV, $\sigma(e^+ e^- \rightarrow \neut_2 \neut_2) < 0.5$ pb for typical
parameter choices, although for $M(\neut_2) \lsi 45$ GeV the cross
section (including enhancement from initial state radiation) is $\gsi
1$ pb.  Thus a small number of events might be present in the 133 GeV
multi-jet sample for each experiment.  Such events should exhibit $\Delta
Q$ consistent with zero.  Production of $\neut_2 \neut_3$ is at least an 
order of magnitude lower. 

A $p \bar{p}$ collider is sensitive to both $\neut_i \ch_j$ production
via $W^*$, and $\neut_i \neut_j$ or $\ch_i \ch_i$ production via
$\gamma^*$ and $Z^*$.  The former has the largest cross section so is
of greatest interest.  A small excess production of $6j$ events is
not a promising signature at a hadron collider so it is fortunate that
there are two possible exceptions to the hadronic decay of {\it inos}.
\begin{enumerate} \item If $tan \beta$ is near 1, i.e., $\mu \approx 70$
GeV, $\neut_3$ has practically no gaugino component and is essentially
a symmetric $\tilde{h}_U+\tilde{h}_D$ state.  Therefore for some range
of parameters, $\neut_3 \rightarrow \gamma \neut_1$ competes with the hadronic
decay $\neut_3 \rightarrow b \bar{b} \gl$.  In this range, $\neut_3 \ch_{1,2}$
gives events with a single $\gamma$, large missing energy, and 3 jets
(the decay products of $\ch_1$ or $\ch_2$).  If $\ch_2 \rightarrow \st_1 + b$
is kinematically possible, $\ch_2 \neut_3$ production will give events
with a single $\gamma$, missing energy, and $b+c + \gl$ jets
reconstructing to the heavier chargino mass, which is predicted to lie
in the range 110-120 GeV.  
\item If all squarks are more massive than $\ch_2$, then $\ch_2
\rightarrow W^{\pm} \pho$ will compete with its 3-body hadronic decay.
In this case $\ch_2 \neut_{2,3,4} $ production would give events with
$W, ~E_{miss}$ and 3-jets (or $W, \gamma$ and $E_{miss}$ if $\neut_3
\rightarrow \gamma \pho$).  $\ch_2 \ch_2$ would give events with $W^+ W^- $
and $E_{miss}$.  
\end{enumerate}  

The CDF $e^+e^-\gamma\gamma + E_{miss}$ event\cite{cdf:eegg} could in
principle arise in this scenario, through $\se \se$ production, with
$\se \rightarrow e \neut_3$ followed by $\neut_3 \rightarrow \gamma \pho$.
However a lone $ee\gamma\gamma$ would be very improbable, since the
large higgsino component needed to make $\neut_3 \rightarrow \gamma \pho$
a dominant decay channel reduces its relative production rate in selectron
decay.  

To summarize:
\begin{itemize} 
\item The generic final state for neutralinos and charginos in the
light gaugino scenario is three jets, unless $m(\sneu)$ is small enough
that two-body leptonic decays are allowed.
\item Chargino production and decay gives a good description of the
rate and characteristics of ``$4j$'' events seen by ALEPH at 133
GeV, in the light gaugino scenario (with no R-parity violation), for
chargino mass of $\approx 53$ GeV.  
\item If this is the origin of the anomalous events, the
chargino is probably lighter than any squark, but has a mass comparable
to the electron sneutrino.
\item Independent of the chargino interpretation of the ALEPH $4j$ events
at 133 GeV, neutralino masses are $m(\chi^0_1) \sim 1 $ GeV and $m(\chi^0_2)
\gsi 38 $ GeV in this scenario.   
\item If the ALEPH hint of charginos at 53 GeV disappears with more 
statistics, the prospects for finding charginos in the light gaugino
scenario are still good.  At E=190 GeV the cross section for
chargino pair production is greater than 1.4 pb for $M_{\sneu_e} = 100$ GeV,
even for the most pessimistic case of degenerate charginos with mass
$m_W$; for large $M_{\sneu_e}$ the cross section is greater than 4 pb.
The $6j$ signal may be difficult to discriminate from the background, however.

\end{itemize}

{\bf Acknowledgements:}  I am indebted to many people for information
and helpful discussions, including H. Baer, J. Berryhill, J. Carr, J.
Conway, H. Frisch, E. Gross, G. Kane, P. Janot, M. Mangano, F. Richard, M.
Schmitt, A. Tilquin, and S. L. Wu.  



\end{document}